\documentclass[pra, aps, superscriptaddress, twocolumn]{revtex4-1}
\usepackage{hyperref}

\usepackage[utf8]{inputenc}
\usepackage{lmodern}
\usepackage[version=3]{mhchem}
\usepackage{amsmath,amssymb,amstext,amsfonts}
\usepackage{natbib}
\usepackage{psfrag,graphicx}
\usepackage[]{units}
\usepackage{upgreek}
\usepackage{textcomp} 
\usepackage{color}
\usepackage{xcolor}
\usepackage{float}
\usepackage{subfigure}
\usepackage{array} % Matrizen in mathematischen Formeln
\usepackage{multirow}
\usepackage{epstopdf}
\usepackage[ngerman, num]{isodate}

\begin{document}
% ------
% Maketitle metadata
%\title{ Selective Laser Melted and Diffusion Processed NdFeB Magnets to Increase the Coercivity}
\title{Polymer-bonded anisotropic SrFe$_\text{12}$O$_\text{19}$ filaments for fused filament fabrication}

\author{Christian~Huber}
\thanks{Correspondence to: \href{mailto:huber-c@univie.ac.at}{huber-c@univie.ac.at}}
\affiliation{Physics of Functional Materials, University of Vienna, 1090 Vienna, Austria}
\affiliation{Christian Doppler Laboratory for Advanced Magnetic Sensing and Materials, 1090 Vienna, Austria}

\author{Santiago~Cano}
\affiliation{Institute of Polymer Processing, Montanuniversitaet Leoben,  8700 Leoben, Austria}

\author{Iulian~Teliban}
\affiliation{Magnetfabrik Bonn GmbH, 53119 Bonn, Germany}

\author{Stephan~Schuschnigg}
\affiliation{Institute of Polymer Processing, Montanuniversitaet Leoben,  8700 Leoben, Austria}

\author{Martin~Groenefeld}
\affiliation{Magnetfabrik Bonn GmbH, 53119 Bonn, Germany}

\author{Dieter~Suess}
\affiliation{Physics of Functional Materials, University of Vienna, 1090 Vienna, Austria}
\affiliation{Christian Doppler Laboratory for Advanced Magnetic Sensing and Materials, 1090 Vienna, Austria}

%\date{16.11.2016}
\date{\today}

%$ $\newline
%%%%%%%%%%%%%%%%%%%%%%%%

\begin{abstract}
In this publication we describe the extrusion process and the properties of polymer-bonded anisotropic SrFe$_\text{12}$O$_\text{19}$ filaments for fused filament fabrication (FFF). Highly filled polyamide 12 filaments with a filling fraction from 40~vol.\% to 55~vol.\% are mixed and extruded into filaments with a diameter of 1.75~mm. Such filaments are processable with a conventional FFF 3D printer. No modifications of the 3D printer are necessary. Detailed mechanical and magnetic investigations of printed samples are performed and discussed. In the presence of an external alignment field, the Sr ferrite particles inside the PA12 matrix can be aligned along an external magnetic field. The remanence can be increased  by 40~\% by printing anisotropic structures. For the 55~vol.\% filled filament, a remanence of  212.8~mT and a coercivity of 307.4~mT are measured. The capabilities of printing magnetic anisotropic structures in a complex external field are presented with a Halbach-array arrangement. By the aim of an inverse field model, based on a finite element method, the orientation of the particles and the quality of the print can be estimated by a nondestructive method.  
\end{abstract}

\maketitle

\section{Introduction}
Fused filament fabrication (FFF) or fused deposition modeling (FDM) is a well-known and widely used additive manufacturing (3D printing) process that uses a thermoplastic wire-shaped filament to build a workpiece layer-by-layer \cite{3d-print}. By mixing magnetic particles with a thermoplastic matrix material, FFF can be also used to print polymer-bonded hard or soft magnets \cite{pub_16_1_apl, pub_17_1, huber2017topology, ortner2017application, baam, von20183d, khatri20183d, patton2019manipulating}. Most recent publications about FFF of hard magnetic materials use isotropic NdFeB powder for the filament fabrication. NdFeB has the highest maximum energy product $(BH)_\text{max}$ of all commercially available magnetic materials, but due to the large amount of rare-earth materials, it is also one of the most expensive magnet. Another disadvantage is the low working temperature compared to other materials and the negative temperature coefficient of the coercivity \cite{ndfeb}.

$(BH)_\text{max}$ in polymer-bonded magnets is proportional to the volumetric filling fraction of the magnetic powder, and it is barely half of sintered magnets \cite{recent_devel}. According, to the Stoner–Wohlfarth model of single-domain ferromagnets, the maximum remanence $B_r$ of an isotropic magnet is only half of an anisotropic magnet \cite{stoner}. This means, by using anisotropic magnetic materials, $(BH)_\text{max}$ can be increased significantly. However, the alignment of the magnetic easy-axis of the particles along an external field is the challenging task for the manufacturing of anisotropic magnets. 

Anisotropic NdFeB powder can be produced by a hot press (die-upsetting) procedure \cite{die-upsetting}. Magnetic fields larger than 1.2~T are necessary to align anisotropic NdFeB powder particles inside a plasticized thermoplastic matrix \cite{nlebedim2017studies, gandha2018additive}. In this publication, we are using ferrite powder to extrude polymer-bonded filaments for a FFF 3D printer. Barium  (BaO$\cdot$6Fe$_2$O$_3$) or strontium ferrite (SrO$\cdot$6Fe$_2$O$_3$) have a low-price compared to rare-earth permanent magnet materials. They are made by mixing of the barium or strontium carbonate with Fe$_2$O$_3$ at a temperature of about $1200$~$^\circ$C. This material is then ball milled to reduce the particle size, pressed in a die, and sintered at about $1200$~$^\circ$C to make the final magnet \cite{intr_mag_material, ferrite}. To align aniostropic Sr ferrites inside a polyamide matrix, an alignment field $>200$~mT is necessary \cite{klaus}. 

Thermoplastic matrix plastics must have a low processing viscosity and high strength, and the matrix should also provide a high elasticity. Polyamide has a good combination of these properties, and is therefore suitable for the processing of highly filled plastics. Especially polyamides such as PA6, PA11, and PA12 are commercially relevant \cite{diss_drummer, abs}. 

We present the manufacturing process of highly filled polymer filaments with different amounts of SrFe$_\text{12}$O$_\text{19}$ powder. The mechanical and magnetic anisotropic behavior of printed structures are analyzed and characterized in detail. Even more, structures with a complex magnetization distribution are printed and characterized by an inverse field computational model.

\section{Filament extrusion}
Up until now, no commercial filaments made of hard magnetic compound are available. Nevertheless, several commercial hard magnetic compound formulations exist for the manufacturing of polymer-bonded magnets by injection molding. Such feedstock materials can be used to extrude filaments by an extruder \cite{pub_16_1_apl}. However, to keep the full control of the material formulation and the production process, a compounder or a mixing extruder can be used to mix the pure matrix thermoplastic with magnetic particles. 

\begin{figure}[ht]
	\centering
	\includegraphics[width=1\linewidth]{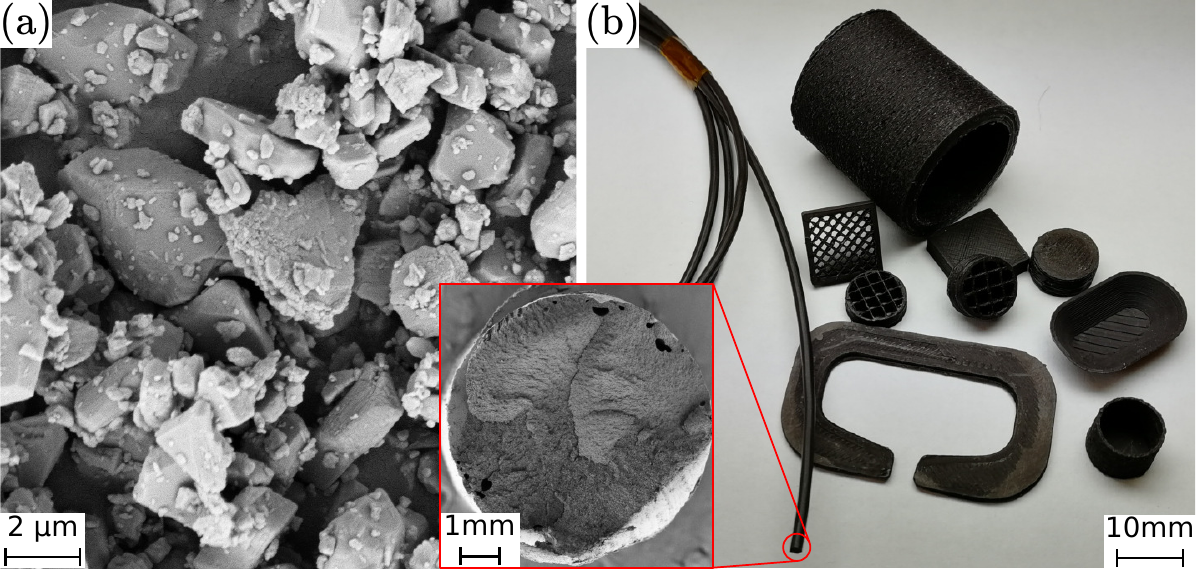}
	\caption{Magnetic filaments for FFF. (a) SEM image of the anisotropic Sr ferrite powder. (b) Picture of the extruded 50~vol.\% filament and several 3D printed samples.}
	\label{fig:filament}
\end{figure}

In this paper, the feedstrock is a compound of PA12 (Grilamid L20G, EMS-Grivory AG) and the magnetically anisotropic  SrFe$_\text{12}$O$_\text{19}$ powder (OP~71, Dowa Electronics Materials Co.,LTD.) with filler fractions of $40$, $45$, $50$, and $55$~vol.\%. Fig.~\ref{fig:filament}(a) shows a  scanning electron microscopy (SEM) image of the powder. The powder has an average particle size of 1.25~\textmu m.

The feedstocks with the different amount of magnetic materials are compounded in a co-rotating twin-screw extruder which is specially designed for compounding of highly-filled polymers with metals or ceramics (Leistritz Extrusionstechnik GmbH, ZSE 18 HPe-48D). The heating zones of the twin-screw are at constant controlled temperatures. The feed section is the coolest with $80$\,$^\circ$C, and the temperature increases up to the die, which has a temperature of $260$~$^\circ$C. Screw rotation is set at $900$~rpm. The extruded compound is pulled away from the die with a conveyor belt  and later granulated in a cutting mill (both Reduction Engineering Scheer). 

\begin{figure}[ht]
	\centering
	\includegraphics[width=1\linewidth]{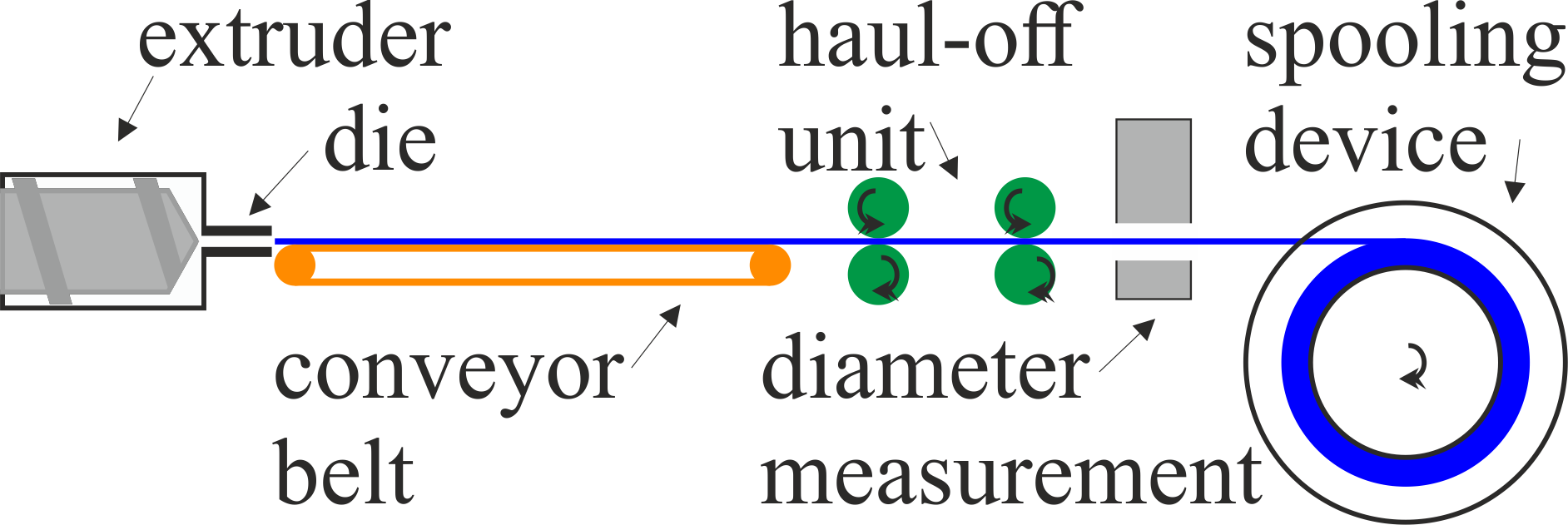}
	\caption{Sketch of the filament extrusion line.}
	\label{fig:extrusion_line}
\end{figure}

To prepare the filaments with a diameter of $1.75$~mm a single screw extruder (Dr. Collin GmbH, FT-E20T-MP-IS) is used. 
%The extruder barrel has five heating zones and they are set at $180/200/205/205/210$~$^\circ$C. The die is also heated to a temperature of $220$~$^\circ$C. The rotational speed of the screw is set at $90$~rpm. 
At the outlet of the extruder die, a polytetrafluoroethylene (PTFE) conveyor belt (Geppert-Band GmbH) is placed to pull the filament as it is extruded. After the filament passes the conveyor belt, it is guided to a haul-off unit. Finally, the filament is winded into spools using a spooling device made by the Montanuniversitaet Leoben. The diameter and ovality of the produced filament are controlled by a diameter-measuring system (SIKORA AG, Laser 2010 T). According to the measured diameter of the filament, the haul-off and the spooling speeds are manually regulated to obtain a filament with an appropriate diameter of $1.75\pm0.2$~mm. A schematic sketch of the filament extrusion line is shown in Fig.~\ref{fig:extrusion_line}.

\section{Material characterization}
The extruded filaments are processable with many FFF 3D printers. A 3D printer extruder for filaments with a diameter of 1.75~mm and a minimum nozzle temperature of about 260~$^\circ$C is necessary. In our case, we are using a Velleman K8200 FFF printer and an e3D Titan Aero extruder for the test prints. Tab.\ref{tab:printer_parameter} shows the best empirically found printer parameters for our setup. The nozzle diameter has a huge impact on the quality of the prints and the minimum feature size of a printable structure. For the highly-filled Sr ferrite filaments, a minimum nozzle diameter of 0.3~mm is necessary. This means, geometrical features with a minimum size of 0.3~mm are printable. 
\begin{table}[ht]
\caption{Best empirically found printer parameters for the highly-filled filaments.}
\label{tab:printer_parameter}
\begin{tabular}{l||l}
\textbf{parameter} & value                \\ \hline \hline
extruder temp. ($^\circ$C)     & $260$                \\
layer height (mm)              & $0.1$                \\
printer speed (mm/s)           & $20$                 \\
fill density (\%)              & $100$                \\
fill pattern                   & rectilinear, contour \\
\multirow{2}{*}{build platform}  & layer of Polyvinyl   \\
                               & acetate (PVA)        \\
bed temp. ($^\circ$C)          & $60$                
\end{tabular}
\end{table}

\begin{figure}[ht]
	\centering
	\includegraphics[width=1\linewidth]{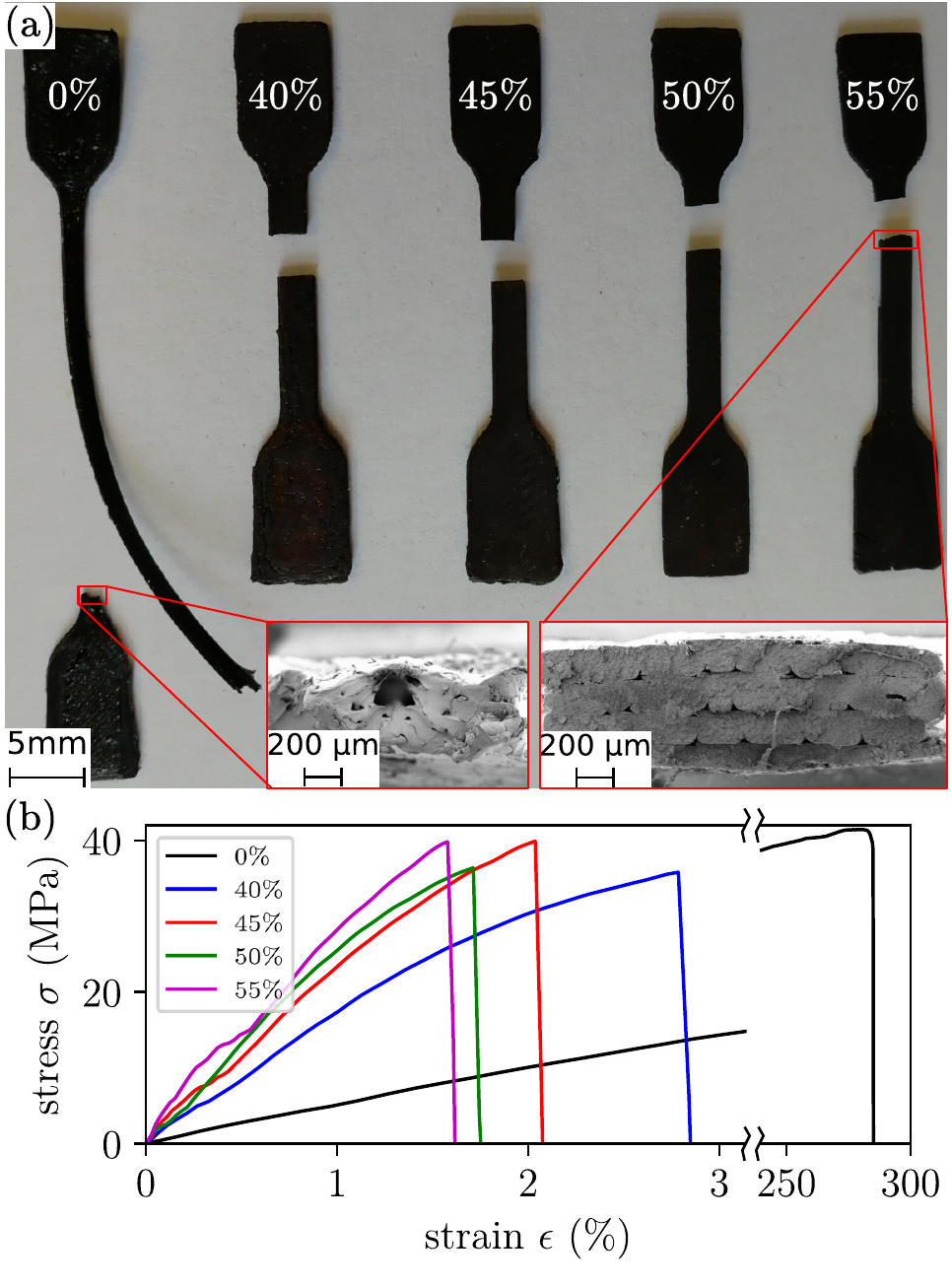}
	\caption{Tensile test results. (a) Picture of printed ISO~527 type~5B specimens after the tensile test. The insert shows a SEM image of the fractured surface for filling fractions of $0$~vol.\% and $55$~vol.\%. (b) Averaged stress-strain curve of the tensile test for the prints with different filling fractions.}
	\label{fig:tensile_test}
\end{figure}

To determine the exact filling fraction $w_f$ inside the polymer, thermal gravimetric analysis (TA-Instruments, TGA~2050) of the materials are performed. Volumetric mass density of printed samples $\varrho_\text{print}$ (printing parameter accordingly to Tab.~\ref{tab:printer_parameter}) measured with a hydrostatic balance (Mettler Toledo, AG204DR) based on the Archimedes' principle. The theoretical volumetric mass density $\varrho_\text{theory}$ is calculated with the densities of PA12 ($\varrho_\text{PA12}=1.01$~g/cm$^3$) and SrFe$_\text{12}$O ($\varrho_\text{SrFe$_\text{12}$O}=5.1$~g/cm$^3$). Tab.~\ref{tab:density} lists the physical properties of the filaments and printed samples with the different filling fractions. Measured powder loads are in good agreement with the specified filling fractions. The densities of the 3D printed filled compounds are around 10~\% lower compared to the theoretical value. This lower density is a result of the FFF printing process itself. The layers of the structure are only placed on top of each other and are not completly compressed. Therefore, air pockets between the layers are hard to be avoided (see Fig.~\ref{fig:tensile_test}(a)). 
\begin{table}[ht]
\caption{Theoretical ($\varrho_\text{theory}$) and printed ($\varrho_\text{print}$) volumetric mass densities, and the measured filling fractions ($w_f$) for the different filaments.}
\label{tab:density}
\begin{tabular}{r||l|l|l|l}
\multirow{2}{*}{\textbf{sample}} & \multirow{2}{*}{\begin{tabular}[c]{@{}l@{}}$\varrho_\text{theory}$\\ (g/cm$^3$)\end{tabular}} & \multicolumn{2}{l|}{$w_f$} & \multirow{2}{*}{\begin{tabular}[c]{@{}l@{}}$\varrho_\text{print}$\\ (g/cm$^3$)\end{tabular}} \\
                        &                                                                                        & (wt.\%)           & (vol.\%)          &                                                                                    \\ \hline \hline
0~vol.\%                & 1.01                                                                                    & --              & --             & 1.000                                                                              \\                        
40~vol.\%                & 2.5                                                                                    & 77.4              & 40.51             & 2.286                                                                              \\
45~vol.\%                & 2.67                                                                                   & 80.2              & 44.61             & 2.440                                                                              \\
50~vol.\%                & 2.84                                                                                   & 83.5              & 50.15             & 2.640                                                                              \\
55~vol.\%                & 3.3                                                                                    & 86.55             & 56.02             & 3.044                                                                             
\end{tabular}
\end{table}

Mechanical properties of the printed work piece are of crucial importance for the design and functionality. For this reason, tensile tests of pure PA12 and the Sr ferrite filaments with different filling fractions are performed. The standard test method for tensile properties of plastics is defined in EN~ISO~527-1. The standard type-5B specimen is printed (printing parameter as listed in \ref{tab:printer_parameter}) for all configurations. This dog-bone shaped specimen has a rectangular cross-section, $1$~mm thick and $2$~mm wide. The total length of the specimen is $35$~mm. The gauge length of the test section is $12$~mm \cite{iso2018}. To reduce the statistical uncertainty and to determine the degree of variability in microstructure and mechanical properties between samples, four test structures are printed and tested. Pictures of the printed specimens after testing for each compound are shown in Fig. \ref{fig:tensile_test}(a).

An universal test machine (RM 100, Schenck Trebel) is used for tensile tests. All tests are performed at room temperature ($295$~K). A plot of the tensile stress-strain curves are shown in Fig.~\ref{fig:tensile_test}(b). It is clearly visible that the tensile strain at break is much higher for pure PA12.

\begin{figure*}[ht]
	\centering
	\includegraphics[width=1\linewidth]{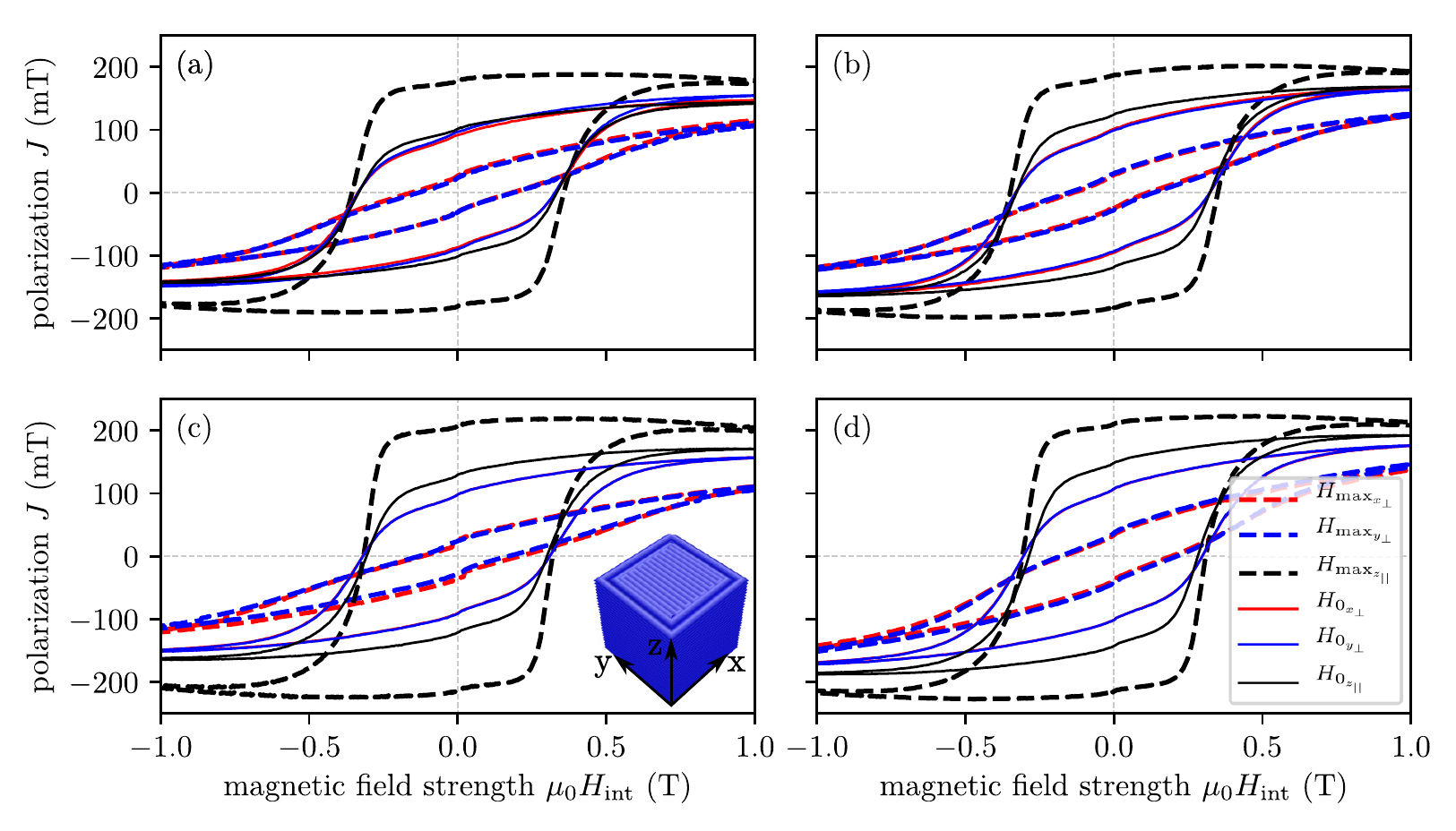}
	\caption{Hysteresis measurements for isotropic (no external field $H_0$) and anisotropic (maximum external field $H_\text{max}$) printed magnets in all magnetization directions. Filling fraction: (a) $40$~vol.\%, (b) $45$~vol.\%, (c) $50$~vol.\%, and (d) $55$~vol.\%.}
	\label{fig:hysteresis}
\end{figure*}

The mechanical properties of highly filled polymer compounds significantly depend on the powder fraction as well as the morphology of the powder particles. Polymer-bonded magnets made of irregular powder shapes have a higher tensile strength compared to those made from atomized spherical powders \cite{mechanical_bonded}.

Test results of tensile tests are summarized in Tab.~\ref{tab:tensile_test}. Manufacturer's data for tensile strength $\sigma_m$ of pure PA12 is $\sigma_m=40$~MPa and a Young's modulus of $E=1100$~MPa.  It should noted, that the tool path, as well as the fill density of the FFF printing process has a crucial impact on the tensile strength of the produced object \cite{ahn2002anisotropic}.
\begin{table}[ht]
\caption{Tensile test results of the PA12 printing filaments with different Sr ferrite filler content. $\sigma_m\ldots$tensile strength, $E\ldots$Young's modulus, and $\epsilon_b\ldots$tensile strain at break.}
\label{tab:tensile_test}
%\resizebox{\textwidth}{!}{%
\begin{tabular}{r||lll}
\textbf{sample}           & $\sigma_m$ (MPa)                  & $E$ (MPa)                         & $\epsilon_b$ ($\%$) \\ \hline \hline
0~vol.\%                & 31.5                                                                                    & 569             & 285                                                                         \\                        
40~vol.\%                & 35.8                                                                                    & 2175            & 2.29                                                                              \\
45~vol.\%                & 39.8                                                                                & 2949            & 2.44                                                                              \\
50~vol.\%                & 36.4                                                                                 & 3298             & 1.75                                                                            \\
55~vol.\%                & 39.9                                                                                   & 4516             & 1.62                            
\end{tabular}
%}
\end{table}

This publication focuses on the magnetic properties of filaments filled with magnetically anisotropic Sr ferrites. For the magnetic hysteresis measurements, cubic samples with a size of 5~mm are printed: (i) without an external alignment field to print magnetic isotropic magnets and (ii) in presence of an alignment field $> 200$~mT to print magnetic anisotropic magnets where the field is high enough to orientate the magnetic easy axis of the ferite particles along the external alignment field \cite{klaus}. All prints are performed with the printing parameters as listed in Tab.~\ref{tab:printer_parameter}. For the anisotropic prints, the cubes are directly printed on the surface of a NdFeB magnet (grade: N40) with a size of $50.8\times50.8\times50.8$~mm$^3$ (L$\times$W$\times$H). This magnet generates a magnetic field of around 550~mT at the surface.

For the measurement of the magnetic hysteresis curve and the magnetic properties of the samples, a permagraph (magnetic closed loop measurement) from Magnet-Physik Dr.~Steingroever GmbH with a JH~15-1 pick-up coil is used. The magnetic hysteresis curved are shown in Fig.~\ref{fig:hysteresis}, and the magnetic properties are summarized in Tab.~\ref{tab:mag_properties}.
\begin{table*}[ht]
\caption{Summary of the magnetic properties for printed samples with different filling fractions in an external field (max, anisotropic) and without an external field ($0$~T, isotropic).}
\label{tab:mag_properties}
\begin{tabular}{ll||lll|lll}
\textbf{sample}                    & ext. field           & $B_{r_z}$ (mT)  & $B_{r_y}$  (mT) & $B_{r_x}$  (mT) & $H_{cj_z}$  (mT) & $H_{cj_y}$  (mT) & $H_{cj_x}$  (mT)       \\ \hline \hline
\multirow{2}{*}{40~vol.\%} & max & 178.2     & 23.7      & 26.8      & 355.8     & 134.0     & 157.0 \\
                          & $0$~T  & 101.8     & 96.7      & 92.3      & 333.7     & 336.0     & 336.3 \\ \hline
\multirow{2}{*}{45~vol.\%} & max & 187.2     & 30.5      & 28.7      & 349.2     & 155.3     & 136.8 \\
                          & $0$~T   & 124.7     & 99.3      & 101.6     & 328.0     & 329.9     & 330.6 \\ \hline
\multirow{2}{*}{50~vol.\%} & max & 207.9     & 24.2      & 26.5      & 316.8     & 132.2     & 121.3 \\
                          & $0$~T   & 129.7     & 98.2      & 97.6      & 306.7     & 319.0     & 320.4 \\ \hline
\multirow{2}{*}{55~vol.\%} & max & 212.8     & 36.9      & 33.7      & 307.4     & 183.7     & 171.0 \\
                          & $0$~T   & 149.0     & 107.6     & 107.2     & 286.0     & 304.1     & 304.9
\end{tabular}
\end{table*}

All samples are measured in all three directions. The $z$-axis is the building direction of the printed cubes as well as the the direction of the alignment field for the anisotropic printed structures ($H_\text{max}$). For the isotropic case ($H_0$, solid lines in Fig.~\ref{fig:hysteresis}), the hysteresis loop and the remanence $B_r$ is almost identical for all directions for the 40~vol.\% sample (Fig.~\ref{fig:hysteresis}(a)). Interestingly, the remanence along the $z$-axis $B_{r_z}$ increases compared to the $x$ and $y$-axis for higher filling fractions. Normally there are more defects/porosity in the $z$-direction due to the poor bonding between layers, which result in a inhomogeneity of the volumetric mass density. This means that the layer structure of highly-filled prints generates a magnetic anisotropic behavior. Nevertheless, it can be seen that the remanence for all directions increases linearly with the filling fraction.

Samples printed on top of the NdFeB magnets show anisotropic behavior as expected (dashed lines in Fig.~\ref{fig:hysteresis}). The magnetic easy-axis of the ferrite particles are orientated along the $z$-axis. The remanence $B_{r_z}$ is around 40~\% higher compared to prints without an external field but lower compared to the theoretical maximum of 50~\% \cite{stoner}.  

In comparison, FFF prints with isotropic NdFeB powder (MQP-S, Magnequench) with a filling fraction of 55~vol.\% have a remanence of $B_r=344$~mT and a coercivity of $\mu_0H_{cj}=918$~mT \cite{pub_16_1_apl}. Nevertheless, polymer-bonded ferrite magnets have some benefits compared to NdFeB bonded magnets. This includes: (i) anisotropic structures can be printed in the presence of magnetic fields $>200$~mT, (ii) lower saturation magnetization is necessary, (iii) no oxidation of the powder, (iv) no rare-earth materials, and (v) significantly lower price.

\section{Inverse field modeling}

\begin{figure*}[ht]
	\centering
	\includegraphics[width=1\linewidth]{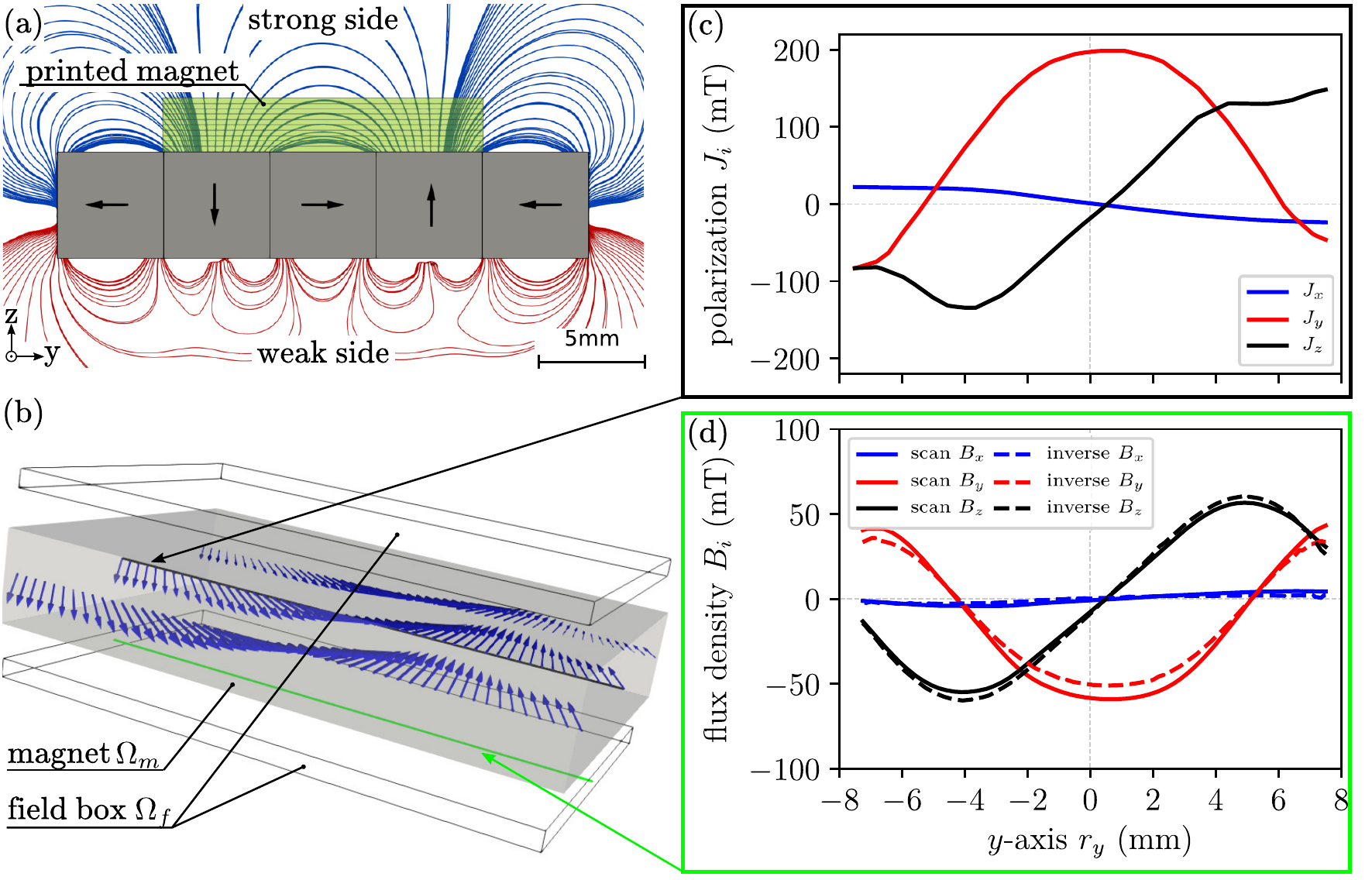}
	\caption{Inverse field modeling of an 3D printed sample on the strong side of a Halbach-array. (a) FEM simulation of the magnetic stray field of a Halbach-array. (b) Reconstructed magnetization inside the magnet $\Omega_m$. The input of the inverse field model uses the magnetic field scans in the field boxes $\Omega_f$. (c) Reconstructed polarization $\boldsymbol{J}$ in the middle of the magnet, along the $y$-axis. (d) Comparison of the measured magnetic flux density $\boldsymbol{B}$ and the simulated $\boldsymbol{B}$ for the reconstructed polarization, $1.25$~mm under the magnet and along the $y$-axis.}
	\label{fig:halbach}
\end{figure*}

This section investigates the possibility of additive manufactured and magnetically anisotropic structures, and how we can examine the quality of the anisotropy by means of a simulation technique. A cuboid magnet with a size of $15\times10\times2.5$~mm$^3$ (L$\times$W$\times$H) is printed with the 40~vol.\% Sr ferrite filament under an external magnetic alignment field. In our case, a permanent magnetic NdFeB (grade: N42) Halbach-array generates the external field \cite{mallinson1973one}. The Halbach-array is assembled by five magnets with a size of $5\times20\times5$~mm$^3$ (L$\times$W$\times$H). Fig.~\ref{fig:halbach}(a) shows a sketch of the Halbach-array and the printed magnet. A finite element method (FEM) simulation shows the magnetic field lines of the Halbach-array. This magnetic arrangement produces a maximum field on one side, and a minimum field on the other side of the structure. During the printing process, the anisotropic particles are aligned along the magnetic field lines of the Halbach-array.

The magnetization $\boldsymbol{M}$ or polarization $\boldsymbol{J}=\mu_0\boldsymbol{M}$ distribution inside a magnet cannot be measured by a nondestructive method. However, by the means of an inverse stray field computation method, the polarization can be reconstructed from magnetic flux density measurements outside the magnet \cite{pub_17_1, inverse_flo}. After the printing process of our magnet, the stray field above and under the magnet in a distance of 1~mm is measured with a 3D magnetic field scanner \cite{pub_16_1_apl}. Each field box $\Omega_f$ has a size of $15\times10\times0.5$~mm$^3$ (L$\times$W$\times$H) and consists of 7200 measurement points of the magnetic flux density vector $\boldsymbol{B}$. Fig.~\ref{fig:halbach}(b) shows the model of the magnet $\Omega_m$ and the field boxes $\Omega_f$.

It exists several well-established simulation methods to calculate the stray field distribution of a magnetic region with a given magnetization distribution \cite{stray_field_fem}. This so-called forward simulation model is well defined and can be calculated by FEM methods and a truncation method to approximate the open boundary condition. In comparison to the forward problem, the inverse problem calculates the polarization for a given stray field distribution. This inverse problem is ill-posed, this means that additional information is necessary to find reasonable results \cite{inverse}. Here, the Tikhonov regularization is implemented in the inverse-stray field computation framework. Solving the minimization problem of the following objective function ($\min_{\boldsymbol{J}}F$), results in the unknown polarization $\boldsymbol{J}$ for each finite element of the model in the region $\Omega_m$
\begin{align}
\label{eq:min}
 %\min_{\boldsymbol{M}}\Arrowvert \boldsymbol{h_{sim}} - \boldsymbol{h_{exp}}\Arrowvert^2_2 + \underbrace{\alpha\Arrowvert \nabla \boldsymbol{M}\Arrowvert^2_2}_{\mathrm{regularization}}
   F = \int_{\Omega_{f}} | \boldsymbol{H_{\text{sim}}}-\boldsymbol{H_{\text{scan}}}|^2 \mathrm{d} \boldsymbol{r} + \underbrace{\alpha\int_{\Omega_{m}}| \nabla \boldsymbol{J}|^2 \mathrm{d} \boldsymbol{r}}_{\mathrm{regularization}}
\end{align}
where $\boldsymbol{H_{\text{sim}}}$ is the stray field calculated by the forward problem in the defined region $\Omega_f$.  $\boldsymbol{H_{\text{scan}}}$ is the measured stray field in the same region $\Omega_f$. $\alpha\geqslant 0$ is the Tikhonov regularization parameter. In this case, $\alpha$ has unit m$^2$. Usually, a smooth polarization distribution is desired, this leads to a minimization of $\nabla \boldsymbol{J}$ in $\Omega_m$.

The main challenge for this regularization is the proper choice of a suitable parameter $\alpha$. If $\alpha$ is too small, the solution will be dominated by the contributions from the data errors. If $\alpha$ is too large, the solution is a poor approximation of the original problem. A well-known method to find an optimal $\alpha$, is the so-called L-curve method \cite{hansen1993use}. A Tikhonov regularization parameter of $\alpha=10^{-3}$~mm$^2$ provides a good results for our example.

The forward and inverse problem simulation model is implemented in a FEM method based on the open-source package Firedrake and Dolfin-adjoint \cite{rathgeber2017firedrake, farrell2013automated}. 

For the measured field distribution, the model gives a polarization distribution inside the magnet $\Omega_m$ as plotted in Fig.~\ref{fig:halbach}(c). Fig.~\ref{fig:halbach}(d) compares the measured magnetic flux density in the middle of the field box $\Omega_f$ with the result of the forward FEM simulation for the reconstructed polarization. The measured stray field distribution is in good agreement with the reconstructed one. It can be seen in Fig.~\ref{fig:halbach}(c) that the anisotropic particles inside the magnet are full orientated along the external field, because the magnitude of the polarization has the same value as the remanence for the 40~vol.\% filament (Fig.~\ref{fig:hysteresis}(a)).

\section{Conclusion}
FFF has the possibility to create polymer-bonded magnets with a complex shape. However, the function of  magnetic systems can also influenced by the design of the magnetization distribution inside the magnet. Such magnetically anisotropic polymer-bonded magnets are well-known, but for the first time, an additive manufacturing method is used to create structures with a complex magnetization distribution and shape.

Higher filler contents increases the magnetic performance of polymer-bonded magnets but on the other side, the processability is getting worse. In our case, a maximum filling fraction of 55~vol.\% of the anisotropic SrFe$_\text{12}$O$_\text{19}$ inside a PA12 polymer is possible to print. However, during printing the feeding of the filament has to be done in a very controlled way due to the brittle filaments. Good printing results can be realized with filling fractions between 40--50~vol.\%. A higher printing temperature decreases the viscosity of the compound, which results in a better flowability through the printing nozzle. Even more, the orientation of the ferrite particles in an external field requires lower magnetic fields.

The mechanical properties like the Young's modulus and tensile strength of FFF produced objects are lower compared to objects manufactured by injection-molding. Young's modulus and the tensile strength increases for higher filler contents. Nevertheless, there is some optimization potential in order to increase the mechanical properties by printing parameters like: layer height, nozzle diameter, tool path generation, build chamber temperature, and infill. The volumetric mass density is also not in the same range as for conventional processed objects, because voids between the strands can hardly be avoided. Further research should investigate the possibility to increase the density and therefore the maximum energy product of the printed magnet.

The magnetic properties of the produced parts are investigated in detail. Samples printed without an external alignment field have around 40~\% lower remanence compared to samples printed in the presence of an external alignment  field. The remanence increases linearly with the filler content of the filament. For filaments with a ferrite content higher than 45~vol.\%, the layer structure of the prints influences the remanence. This could be a result of an unequal mass distribution inside the printed object. Further investigations are necessary to determine this theory.

To test the alignment of the ferrite particles during the printing process, an inverse field model is described. This model can simulate the magnetization distribution inside the magnet only from stray field measurements outside the magnet. We can see that in the presented example, the printed structure is completely aligned along the field of the permanent magnetic Halbach-array. This method can be used for quality checks of any kind of magnetic systems.

\section*{Acknowledgment}
The support from CD-Laboratory AMSEN (financed by the Austrian Federal Ministry of Economy, Family and Youth, the National Foundation for Research, Technology and Development) is acknowledged. The SEM images and the tensile tests are carried out using facilities at the Faculty Center for Nano Structure Research, University of Vienna, Austria.

\section*{References}
%\bibliographystyle{elsarticle-num-names}
%\bibliographystyle{aipnum4-1}
%\bibliography{bibliography}


\end{document}